\documentclass[12pt]{article}
\usepackage{epsfig}
\begin{document}
\title{The critical perturbation parameter estimate\\
for the transition from  regularity to chaos\\
in quantum systems}
\author{V. E. Bunakov, I. B. Ivanov, R. B. Panin
\footnote{e-mails: Vadim.Bunakov@pobox.spbu.ru, Ivan.Ivanov@thd.pnpi.spb.ru}
\\
Petersburg Nuclear Physics Institute, 188300 Gatchina, Russia\\}
\date{}
\maketitle
\begin{center}
{\bf Abstract}
\end{center}
{\small
In this paper we continue to develop our approach
to the chaoticity properties of the quantum Hamiltonian
systems. Our earlier suggested
chaoticity criterion characterizes the initial
symmetry breaking and the destruction of the
corresponding integrals of motion in a perturbed
system, which causes the system's chaotisation.
Transition from regularity to chaos in quantum systems
occurs at a certain critical value of the perturbation parameter.
In our previous papers we had to diagonalize the
perturbed system's Hamiltonian matrix in order to
estimate this parameter.
In the present paper we demonstrate that the critical perturbation
parameter for the transition from regularity to
chaos  can be estimated in the framework of the
first order perturbation theory.
The values of thus obtained critical parameter
are in good agreement with the results of our
previous precise calculations for Hennon-Heiles
Hamiltonian and diamagnetic Kepler problem.
}

\section{Introduction}

For last decades the investigations of quantum chaos have been extensively
carried out but this field remains under hot discussions. One part
of researchers believe that quantum chaos doesn't exist and at best should
be studied in the semiclassical approximation. It is believed
 that the quantum analogue of the classical system should have properties
reflecting regularity or chaoticity of classical trajectories. The law of
level spacing distribution is considered to be one of such properties.
It is also believed that the quantum analogue of the classically chaotic
system obeys
Wigner level spacing law, while Poissonian law holds for regular systems.
However many authors (including us [1]), pointed out the
incompleteness and crudeness of this criterion of quantum chaoticity.

In this paper we continue to develop our approach [2-6] to the chaotic
properties of the quantum Hamiltonian systems. Our main point is connection
between the symmetry properties of a system and its regularity or chaoticity.
We had demonstrated that our earlier suggested chaoticity parameter characterizes
the
initial symmetry breaking and destruction of the corresponding integrals of
motion in a perturbed system, which leads to chaotisation.
As it was shown in [3], in the semiclassical limit the chaoticity parameter
transforms into Lyapunov's exponent of the corresponding classical
motion. We had also shown that according to our chaoticity criterion
the transition from regularity
to chaos in classical and quantum systems occurs at the same critical
perturbation parameter.

In our previous papers we had to diagonalize the perturbed system's Hamiltonian
matrix in order to estimate this perturbation parameter. In this paper we shall
show that it is possible to estimate it in the framework of the
perturbation theory. In part 2
we reconsider the definition of our chaoticity parameter which is used for
the investigation of the transition from
regularity to chaos .
Part 3 is devoted to the analytical estimate of the critical
parameter in the framework of the first order perturbation theory.
The comparison of our estimate with the precise calculations for Hennon-Heiles
Hamiltonian and diamagnetic Kepler problem is done in part 4.

\section{Destruction of quantum numbers\\
and the chaoticity criterion}

Consider a stationary quantum system with Hamiltonian $H$
as a sum of Hamiltonian $H_0$ of an integrable system and
perturbation $\lambda V$, which destroys the symmetries of $H_0$ in such a way
(see e.g. [3-4]) that the number $M$ of the independent first integrals of
motion becomes less than the number of the system's degrees of freedom:
\begin{equation}
\label{H}
H=H_0+\lambda V.
\end{equation}
Eigenfunctions $\{\phi_\alpha\}$ of the unperturbed integrable
Hamiltonian $H_0$
are common for some complete set of independent mutually commuting
operators $\{J_\rho\}_{\rho=1}^N$ ($N$ - the number of degrees of
freedom of the Hamiltonian $H_0$). Eigenstates $\psi_i$ of the full
Hamiltonian may be expanded in eigenfunctions $\phi_\alpha$
of the unperturbed Hamiltonian $H_0$:
\begin{equation}
\label{expansion}
\psi_i=\sum_\alpha \phi^*_\alpha\psi_i\phi_\alpha =
    \sum_\alpha c^{\alpha}_i\phi_\alpha.
\end{equation}

Let us consider the probability $P_\alpha(E_i)$ to find the basis state
$\phi_\alpha$ in the state $\psi_i$ with energy $E_i$,
which is equal to the squared
absolute value of the corresponding coefficient in (\ref{expansion})
and define the energy width $\Gamma_{spr}^\alpha$ of $P_\alpha(E_i)$ distribution:
the minimal energy interval for which the sum of probabilities
$P_\alpha(E_i)$ is larger or equal to $0.5$.
Thus defined $\Gamma_{spr}^\alpha$ is the energy spreading width of basis
state $\phi_\alpha$. The spectrum of $H_0$ may be degenerate
and then the irreducible representations of the symmetry group of $H_0$
consist of several basis functions which belong to one energy level (shell).
We want to find a parameter characterizing the measure
of initial symmetry breaking of $H_0$ under the influence of
the perturbation $V$. It's clear that
such symmetry breaking results only due to significant mixing between
functions from different irreducible representations. The mixing of
states within one shell doesn't change their symmetry. While the
spreading width $\Gamma_{spr}^\alpha$ is smaller then the distance $D_0$
between the neighboring levels in the spectrum of $H_0$ we can distinguish
"localization domain" (in energy) of one set of basis states from
"localization domain" of another one. When the spreading width exceeds
$D_0$ we start loosing the "traces" of basis functions in the spectrum of
$H$ and can't even approximately compare states $\psi_i$ with irreducible
representation of symmetry group $H_0$. Thus a parameter
\begin{equation}
\label{kappa}
\ae^\alpha = \Gamma_{spr}^\alpha /D_0
\end{equation}
is the natural measure of symmetry breaking.
When the parameter $\ae^\alpha$ exceeds unity the symmetry of the
Hamiltonian $H_0$ disappears. Such a value of perturbation is accompanied by
disappearance of the initial selection rules, the levels are distributed
approximately uniformly (level repulsion) and the level spacing
distribution approaches Wigner's law.
One can say that the transition from regularity to chaoticity
has taken place in quantum system and $\ae^\alpha$
may be considered as the parameter of chaoticity.

The spreading width $\Gamma_{spr}^\alpha$ depends on the number $\alpha$
of the basis state of Hamiltonian $H_0$, i.e. on its quantum numbers.
The classical analogy to this is the dependence of the
invariant torus stability on the corresponding values of integrals of motion.
It is clear that in order to obtain the global chaoticity
characteristic in quantum
case it is necessary to average $\Gamma_{spr}^\alpha$ over the basis states
$\phi_\alpha$ belonging to the same irreducible representation (shell).
Thus we should determine the energy width $\Gamma_{spr}$ of the following
distribution
\begin{equation}
\label{dstr}
P_N(E_i)=
\frac1{dim\mathcal {T}_N}\sum_{\alpha\in\mathcal {T}_N}
|\phi^*_\alpha\psi_i|^2,
\end{equation}
where $\mathcal {T}_N$ is a degenerate level subspace with main
quantum number $N$.
The averaged parameter $\Gamma_{spr}$ unlike the local $\Gamma_{spr}^\alpha$
has an important feature of invariance with respect to the choice of the basis
for the integrable Hamiltonian $H_0$ [1]. From the theoretical point of view the
above chaoticity parameter has one more useful property. As it was shown in
[3], in the semiclassical limit $\Gamma_{spr}^\alpha/\hbar$ transforms into
Lyapunov's exponent of the corresponding classical motion.
Thus we see that parameter $\ae(\lambda ,E)=\Gamma_{spr}/D_0$
"measures" destruction (fragmentation) of the irreducible representations of the
basis and, hence of the Casimir operator (the main quantum number), which is
the approximate integral of motion for small perturbations.

\section{Analytical estimate of the critical\\ perturbation parameter}

The analytical estimate can be performed in the following way:
Firstly, the expansion coefficients (\ref{expansion}) at small parameter
$\lambda$
can be computed in the framework of usual perturbation theory.
Secondly, since it is inconvenient to work with the distribution width
 $P_N(E_i)$ analytically,  we shall use the equivalence of our
chaoticity criterion based on $\ae$ and the criterion of Hose-Taylor approach.

In a series of publications Hose and Taylor (see, e.g. [7,8]) in the framework
of the theory of effective
Hamiltonians suggested the criterion of their existence together
 with their integrals of motion.
According to this criterion, one can construct a convergent sequence
of approximations to the effective hamiltonian provided that a square
projections of a perturbed wave function on model
space is larger than 0.5. Operators commuting with
effective hamiltonian will be approximate integrals of motions of the problem.
Thus, if a square of a projection of some states $\psi_i$ of the Hamiltonian
$H$ on space the $\mathcal {T}_N$ of degenerate the level of $H_0$ are
larger than 0.5, then
the principal quantum number must be an approximate integral of motion for
these states (in this range of energies). Therefore the average
square of a projection of perturbed wave functions on subspace
$\mathcal {T}_N$ with fixed value of main quantum number should reach
value $0.5$ approximately at same critical perturbation
when $\ae=1$ [5].

The principal problem in using the perturbation theory for
estimating of squares of projections is the degeneracy of basis
states and, hence, the necessity to solve a secular equation,
which prevents to obtain the analytical expressions.
In order to bypass this complexity we shall act as follows.
We shall construct such a quantity, which on the one hand will help us
to estimate the critical perturbation parameter, and on the other will be
is invariant with respect to block unitary transformations
inside subspaces $\mathcal {T}_N$. This property
will allow us in evaluating this quantity to utilize not only
the states obtained from the secular equation, but also
any others, generated from them by the block unitary transformations,
(including the initial basis states $\phi_\alpha$).
In this way we shall not neeed to solve the secular equation.

Let us estimate the average square of projection $W(\lambda, N)$ of the
perturbed wave functions not on a subspace $\mathcal {T}_N$, but on its
orthogonal adjoint $\mathcal H\ominus\mathcal {T}_N$.
Further, let us assume, that we already have solved the secular equation
in all subspaces $\mathcal {T}_N$ and found correct functions
$\phi^{(0)}_\alpha $ in the zero approximation.
Then in the first order approximation the expansion coefficients
of perturbed states $i$ on basis $\alpha$ belonging to other shells will be:
\begin{equation}
\label{c}
c_{\alpha i}=\frac {V_{\alpha i}}{E^{(0)}_i-E^{(0)}_\alpha},
\end{equation}
where $V_{\alpha i}=\phi^{(0)*}_\alpha V\phi^{(0)}_i$ and
$E^{(0)}_\alpha$ is the energy of states in a zero approximation.
The square of projection of a perturbed state $\psi_i$ on orthogonal
adjoint to $\mathcal {T}_N$ will be
$$
\sum_{\alpha\notin\mathcal {T}_N}\frac{|V_{\alpha i}|^2}
{(E^{(0)}_i-E^{(0)}_\alpha)^2}.
$$
Now we average the previous expression over all $dim\mathcal {T}_N$
states $\psi_i$ (for small perturbations every state $\psi_i$
can be approximately referred to some $\mathcal {T}_N$):
\begin {equation}
\label {W}
W(\lambda, N)=\frac{1}{dim\mathcal {T}_N}\sum_{i\in\mathcal {T}_N}
\sum_{\alpha\notin\mathcal {T}_N}\frac {|V_{\alpha i}|^2}{(E^{(0)}_i-E^{(0)}_\alpha)^2}.
\end {equation}
With increasing perturbation the parameter $W(\lambda, N)$
will gradually grow (the basis states are fragmented over other shells)
and at some critical value will reach $0.5$.
Thus a required critical perturbation parameter can be found by
solving the equation
\begin{equation}
\label{eq}
W(\lambda, N)=0.5
\end {equation}
with respect to  $\lambda$ or $N$.

Let us prove now the invariance of $W(\lambda, N)$ with respect to
arbitrary block unitary transformations of the basis, which
mixes the functions $\phi^{(0)}_\alpha$ only inside the
irreducible representations $\mathcal {T}_N$ ( the
functions from different $\mathcal {T}_N$ are not mixed).
It is possible to present the sum of the states $\alpha\notin {T}_N$as
\begin {equation}
\sum_{\alpha\notin\mathcal {T}_N}=\sum_{n\ne N}\sum_{\alpha\in\mathcal {T}_n}.
\end {equation}
Therefore ~(\ref {W}) can be written as
\begin {equation}
\label{W1}
W(\lambda, N)=\frac{1}{dim\mathcal {T}_N}\sum_{n\ne N}W_n, \\
\end {equation}
\begin {equation}
\label{Wn}
W_n=\sum_{i\in\mathcal {T}_N}\sum_{\alpha\in\mathcal {T}_n}
\frac{|V_{\alpha i}|^2}{(E^{(0)}_i-E^{(0)}_\alpha)^2}.
\end {equation}
Now consider what happens with the value $W_n$ under the block unitary
transformation $\hat U_b$ of the zero approximation basis wave functions
\begin {equation}
\phi^{(0)}_i\longmapsto\tilde\phi^{(0)}_i=\sum_\mu U_{\mu i}\phi^{(0)}_\mu.
\end {equation}
Since the energies $E^{(0)}_i$ and $E^{(0)}_\alpha$ are the same for
all states $i\in\mathcal {T}_N$ and $\alpha\in\mathcal {T}_n$ and
do not vary under the transformation $\hat U_b$, the denominator in
(\ref {Wn}) we can be taken out of the sum.
Utilizing the symmetry in labels $i$ and $\alpha$ we can prove the invariance
of $W_n$ with respect to their exchange.
Indeed,
$$
(E^{(0)}_i-E^{(0)}_\alpha)^2\tilde W_n=\sum_{i\in\mathcal {T}_N}
\sum_{\alpha\in\mathcal {T}_n}(\phi^{(0)*}_\alpha
V\tilde\phi^{(0)}_i)(\tilde\phi^{(0)*}_i V\phi^{(0)}_\alpha)=
$$
$$
\sum_{i\in\mathcal {T}_N}\sum_{\alpha\in\mathcal
{T}_n}\phi^{(0)*}_\alpha V\sum_\mu U_{\mu i}\phi^{(0)}_\mu\sum_\nu
U_{\nu i}^*(\phi^{(0)*}_\nu V\phi^{(0)}_\alpha)=
\sum_\alpha\phi^{(0)*}_\alpha V\sum_{\mu,\nu}(\sum_i U_{\mu i}
U^*_{\nu i})\phi^{(0)}_\mu(\phi^{(0)*}_\nu V\phi^{(0)}_\alpha)=
$$
$$ \sum_\alpha\phi^{(0)*}_\alpha
V\sum_{\mu}\phi^{(0)}_\mu(\phi^{(0)*}_\mu V\phi^{(0)}_\alpha)=
\sum_\alpha\sum_{\mu}|V_{\alpha\mu}|^2=(E^{(0)}_i-E^{(0)}_\alpha)^2W_n.
$$
Thus, making transformation $\hat U_b$ at first inside the
subspaces $\mathcal {T}_N$, and then inside $\mathcal {T}_n$,
we obtain the invariance of $W _ n$ with respect to $\hat U_b$.
Since each term $W_n$ in the sum (\ref{W1}) is invariant,
all the sum $W(\lambda, N)$ will be also invariant with respect to
unitary transformations.

This property allows us to bypass the necessity to solve the
secular equation. We can use in the evaluation of
$W(\lambda, N)$ in Eq.(\ref{W}) the initial basis functions $\phi_\alpha$
and energy $E^{(0)}_\alpha$, and find the critical perturbation parameter value
from the equation (\ref{eq}).

\section{Two examples}

In this part we shall compare the analytical estimates
of the critical perturbation parameter with the precise calculations for
a nonlinear Hennon-Heiles Hamiltonian and diamagnetic Kepler problem.

\begin{figure}
\epsfxsize = 9,360cm
\epsfysize = 6.867cm
\epsfbox{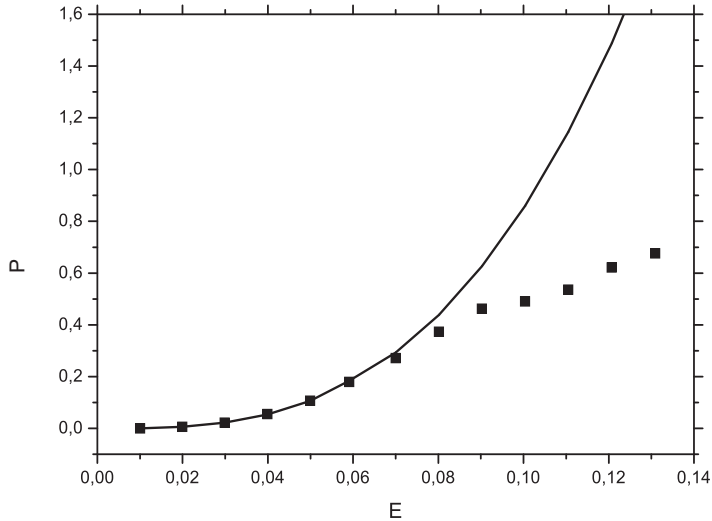}
\caption{Perturbation theory (line) and precise calculation (points) of
average square of a projection of wave functions on a subspace
$\mathcal H\ominus\mathcal {T}_{N} $ dependence as functions of energy in the
Hennon-Heiles system.
\label{fig:pthh}}
\end{figure}

Let us consider a known Hennon-Heiles system with hamiltonian:
\begin{equation}
\label{hhh}
H(q,p)=\frac{1}{2}(p^2_1+q^2_1)+\frac{1}{2}(p^2_2+q^2_2) +
\lambda (q^{2}_1 q_2-q^3_2/3).
\end {equation}
Introducing operators of creation and annihilation
$a^\dagger _k = \frac{1}{\sqrt{2}}(q_k + ip_k),
a_k = \frac{1}{\sqrt{2}}(q_k - ip_k), k = 1,2$ we construct, as usual,
the two-dimensial Cartesian oscillator basis
\begin{equation}
\label{cob}
\phi_\mu = \phi_{n_1n_2} = \frac{1}{\sqrt{n_1!n_2!}}
(a^\dagger _1)^{n_1}(a^\dagger _2)^{n_2}\phi_0,
\end{equation}
$$
E_{n_1, n_2}=E_N=\hbar (n_1 + n_2 + 1).
$$
It is easy to write out matrix elements of the Hamiltonian (\ref {hhh})
in the basis (\ref {cob}) and to find through a diagonalization the
system's eigenergies and eigenstates.
In the calculations of bound states (with energies $E < 1/(6\lambda^2)$)
the value $\lambda = 1$ and $\hbar= 0.01$ was fixed
and the basis (\ref{cob}) was used, which included $496$ of states ($30$ shells).
We investigated average (over states with approximately equal energy)
square of a projection of perturbed wave functions on subspace
$\mathcal H\ominus\mathcal {T}_N$.
First we carried out precise calculations (see points on fig. \ref {fig:pthh}),
and then determined the value $W (\lambda, N)$ from Eq.(\ref {W}) (line on Fig.
\ref {fig:pthh}).
According to precise calculation we achieve critical value $0.5$ at an energy
$E\approx 0.105$, that practically coincides with earlier obtained [4] value with the
criterion $\ae = 1$ ($ E\approx 0.11 $).
As one should to expect, the perturbation theory works well
at small perturbation parameter (up to energies $E\approx 0.08$)
and as a result yields the critical value of energy
$E\approx 0.084$.

\begin{figure}
\epsfxsize = 10.467cm
\epsfysize = 6.867cm
\epsfbox{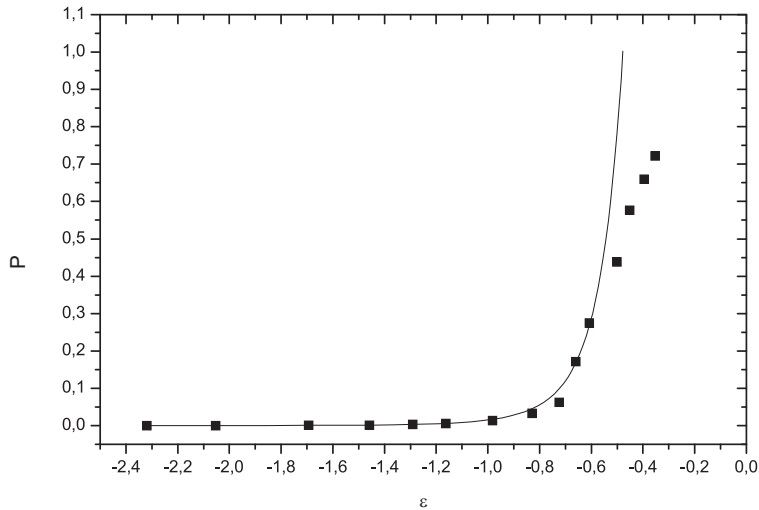}
\caption{Perturbation theory (line) and precise calculation (points) of average
square of a projection of wave functions on a subspace
$\mathcal H\ominus\mathcal {T}_{10}$ dependence as functions of scaled energy
for a diamagnetic Kepler problem.
\label{fig:ptkp}}
\end{figure}

The same test of the critical perturbation parameter estimate
was carried out for diamagnetic Kepler problem  with Hamiltonian
\begin{equation}
H = p^2/2m -e^2/r + \omega l_z + \frac{1}{2} m\omega^2 (x^2 + y^2).
\end{equation}
Here frequency $\omega = eB/2mc$, $B$ --- magnetic field, directional
along an axes $z$. Instead of the usual energy $E$ we used the standard
dimensionless energy $\epsilon = E\gamma^{-2/3} $,
where $\gamma = \hbar\omega / {\cal R}$ (${\cal R}$ --- Rydberg constant).
We diagonalized a matrix of the Hamiltonian in the Coulomb
parabolical basis with quantum numbers $(n_1, n_2, m)$ and
energies of basis states $E_n = -1/2n^2$, $n = n_1 + n_2 + |m| + 1$.
For the investigation of a fragmentation of basis functions with $m = 0$
from a shell with the principal quantum number $10$ we used the basis
of the first $20$ shells. We studied the dependence of the average square of a
projection
of the perturbed wave functions on a subspace $\mathcal H\ominus\mathcal {T}_N$
as a function of the scaled energy.
As in the Hennon-Heiles case, we fixed a subspace
$\mathcal {T}_{10}$ with a main quantum number $N = 10$ and varied
the values $\gamma$. As an averaging range we selected the part of the spectrum,
whose states ($ dim\mathcal {T}_{10}=10$) would have
the maximum square of a projection on $\mathcal {T}_{10}$.
Fig. \ref{fig:ptkp} shows the analytic result $W(\lambda, N)$
(line) and the numerically clculted (points) dependence of the precise wave functions'
average square
projections on a subspace $\mathcal H\ominus\mathcal {T}_{10}$
as functions of the scaled energy $\epsilon$.
As well as in the previous case we have obtained the quite good agreement
of the theoretical critical parameter estimate ($\epsilon_{cr} = -0.54$)
with precise calculation ($\epsilon_{cr} = -0.47$),
which is also in correspondence with earlier results [6] obtained using $\ae=1$ criterion
($\epsilon_{cr} = -0.45$).

\section{Conclusion}

In this paper we continue to develop our approach [2-4] to the chaotic
properties of the quantum Hamiltonian systems. Our main point is the connection
between the symmetry properties of a system and its regularity or chaoticity.
We show that the earlier suggested chaoticity parameter $\ae(\lambda ,E)$
characterizes the
initial symmetry breaking and destruction of the corresponding integrals of
motion in a perturbed system, which leads to chaotisation.
The value of $\ae$ is equal to the ratio of the average
spreading width of the basis wavefunctions over the perturbed ones
to the average unperturbed  level spacings.
In order to estimate the critical
perturbation parameter for the transition from regularity to chaos
we had to diagonalize the perturbed system's Hamiltonian
matrix.
In the present paper we have shown that the critical perturbation
parameter may be estimated in the framework of the first order
perturbation theory. We construct a quantity which is invariant with respect to
unitary transformations of subspaces $\mathcal {T}_N$ of basis degenerated levels $H_0$
and indicate the measure of system chaoticity.
Thus there is no need to deal with secular equation and the approximate value of the
critical
perturbation parameter might be obtained from the
relation which contains matrix elements of perturbation $V$
in the initial basis and energies of basis states.
The values of thus obtained critical parameter
are in good agreement with the results of our
previous precise calculations for Hennon-Heiles
Hamiltonian and diamagnetic Kepler problem.

One of the authors (IBI) is indebted to Prof. Zikiki and to the Organizing
Committee of V. Gribov's Foundation for their support. We also thank
RFFI (grant N 00-15-96610).

\end{document}